\documentclass[journal]{IEEEtran}

\usepackage{graphicx}
\usepackage{multirow}
\usepackage{wrapfig}
\usepackage{subfigure}
\usepackage{amsmath}
\usepackage{cases}
\usepackage{array}
\usepackage{tikz}
\usepackage{tabularx}

\usepackage{algorithmic}
\usepackage{algorithm}
\usepackage{amsfonts}

\ifCLASSINFOpdf
\else
\fi

\begin{document}

\title{Harmonic Detection from Noisy Speech with Auditory Frame Gain\hspace{-0.01cm} for\hspace{-0.01cm} Intelligibility\hspace{-0.01cm} Enhancement\vspace{0.01cm}}

\author{A.~Queiroz,~\IEEEmembership{Student Member,~IEEE,}
        and~R.~Coelho,~\IEEEmembership{Senior~Member,~IEEE}
\thanks{This work was supported in part by the National Council for Scientific and Technological Development (CNPq) 305488/2022-8 and Fundação de Amparo à Pesquisa do Estado do Rio de Janeiro (FAPERJ) under Grant 200518/2023 and in part by the Coordenação de Aperfeiçoamento de Pessoal de Nível Superior - Brasil (CAPES) - under Grant Code 001.}
\thanks{The authors are with the Laboratory of Acoustic Signal Processing, Military Institute of Engineering (IME), Rio de Janeiro, RJ 22290-270, Brazil (e-mail: coelho@ime.eb.br).}
}

\markboth{IEEE/ACM TRANSACTIONS ON AUDIO, SPEECH, AND LANGUAGE PROCESSING}%
{Shell \MakeLowercase{A. Queiroz and R. Coelho}: Harmonic Detection from Noisy Speech with Auditory Frame Gain\hspace{-0.01cm} for\hspace{-0.01cm} Intelligibility\hspace{-0.01cm} Enhancement}

\maketitle

\begin{abstract}

This paper introduces a novel (HDAG - Harmonic Detection for Auditory Gain) method for speech intelligibility enhancement in noisy scenarios. In the proposed scheme, a series of selective Gammachirp filters are adopted to emphasize the harmonic components of speech reducing the masking effects of acoustic noises. The fundamental frequency are estimated by the HHT-Amp technique. Harmonic patterns estimated with low accuracy are detected and adjusted according the FSFFE low/high pitch separation. The central frequencies of the filterbank are defined considering the third octave subbands which are best suited to cover the regions most relevant to intelligibility. Before signal reconstruction, the gammachirp filtered components are amplified by gain factors regulated by FSFFE classification. The proposed HDAG solution and three baseline techniques are examined considering six background noises with four signal-to-noise ratios. Three objective measures are adopted for the evaluation of speech intelligibility and quality. Several experiments are conducted to demonstrate that the proposed scheme achieves better speech intelligibility improvement when compared to the competing approaches. A perceptual listening test is further considered and corroborates with the objective results.

\end{abstract}

\begin{IEEEkeywords}
Gammachirp filtering, low/high frequency separation, harmonic detection, noisy speech.
\end{IEEEkeywords}

\IEEEpeerreviewmaketitle

\vspace{0.15cm}
\section{Introduction}
\vspace{0.25cm}



\IEEEPARstart{A}{coustic} noise is a strong masking effect that impairs speech intelligibility \cite{STEINBERG_1947}\cite{ASSMANN_2004}. This interference underlies several research studies such as speech enhancement \cite{GERKMANN_2012}\cite{TAVARES_2016}\cite{MEDINA_2021}, source localization \cite{DRANKA_2015}\cite{NAYLOR_2018}, robot audition \cite{RASCON_2020}, speech and speaker recognition \cite{LJOLJ_2002}\cite{VENTURINI_2014}. Thus, its mitigation is a relevant element of interest for the intelligibility and quality enhancement. Several signal processing methods are described in the literature to attenuate noise interference for speech quality assessment \cite{FLANDRIN_2014}. However, this achievement not necessarily leads to speech intelligibility improvement \cite{KIM_2011}. On the other hand, acoustic masks \cite{LI_2009}\cite{LOIZOU_2010}\cite{FARIAS_2021} are defined to emulate the \textit{cocktail party} effect. These solutions provide intelligibility enhancement for the target speech signal.


In the last years, the analysis of harmonic components of noisy speech \cite{BROWN_2010}\cite{EALEY_2001} has encouraged the proposal of new strategies for intelligibility gain \cite{FENG_2023}\cite{VASILI_2023}. For these, harmonic components such as fundamental frequency (F0) and formants \cite{NEY_1998} play an interesting role for intelligibility in noisy condition \cite{BROWN_2010}\cite{WANG_2017}\cite{WANG_2018}. Time-domain adaptive solutions are designed to deal with the harmonics of the speech signal to reduce the noise effects. In \cite{GAEL_2016}, the formant center frequencies from voiced segments of speech are shifted away from the region of noise. This formant shifting procedure \cite{GAEL_2017} simulates the human strategy to provide a more audible signal in noisy environment, i.e., the Lombard effect \cite{LOMBARD_1911}. Results showed that the Smoothed Shifting of Formants for Voiced segments (SSFV) is able to improve the intelligibility of speech signals in car noise environment. A different approach was proposed in \cite{QUEIROZ_2021}, where the HHT-Amp \cite{HHT} F0 estimation technique was applied to the harmonic components of noisy speech. The F0-based Gammatone Filtering (GTF$_\text{F0}$) method considered integer multiples of the estimated F0 as center frequencies of a time-domain auditory filterbank. Finally, the outputs are amplified to emphasize the harmonics of the speech signal leading to intelligibility gain.

The use of the Gammatone filterbank in the GTF$_\text{F0}$ method may be limited by the high level masking effects \cite{PATTERSON_1984}. To overcome this issue, the Gammachirp proposed in \cite{PATTERSON_1997} produces a filter with an asymmetric amplitude spectrum. This auditory filter provides an interesting fit to various sets of noise masking data. The center frequencies of the its filterbank must be well-defined considering the relevant ones for intelligibility. In this context, the Extended Short-Time Objective Intelligibility (ESTOI) \cite{JENSEN_2016} performs an evaluation of noisy speech in third-octave subbands. ESTOI also considers the temporal modulation frequencies relevant to speech intelligibility, whose values range from 1--12.5 Hz \cite{PLOMP_1994}\cite{PLOMP_1994_2}\cite{THEUNISSEN_2009}. These subbands and frequency modulation range are able to assist in regulating the bandwidth of filterbanks to cover the harmonic components of speech most relevant for intelligibility.

\begin{figure*}[t!]
\centering
\includegraphics[width=0.99 \linewidth,keepaspectratio=true]{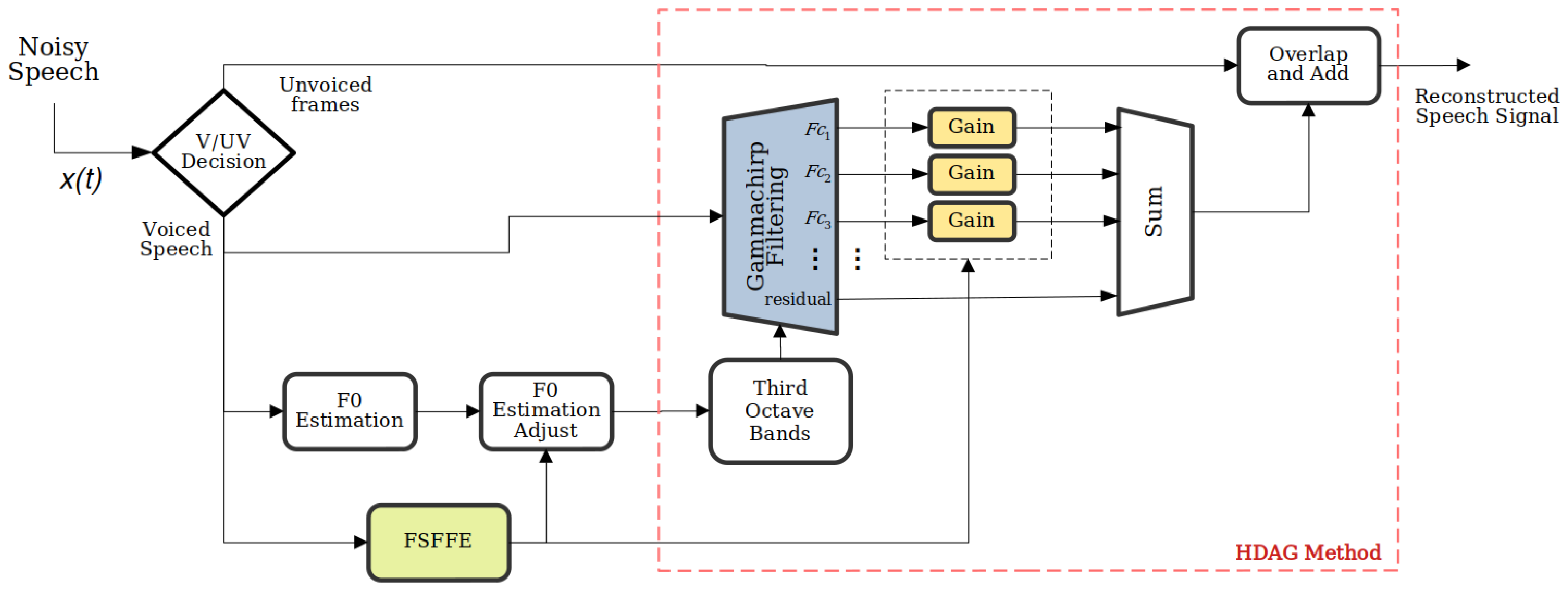}
\caption{Block diagram of the proposed HDAG method for improve the intelligibility of noisy speech signals.}
\label{block_diag}
\end{figure*}

This paper introduces the HDAG (Harmonic Detection with Auditory Gain) method to attain intelligibility enhancement for harmonic components of noisy speech signals. The proposed solution is performed in four steps. Initially, the HHT-Amp method \cite{HHT} is applied to estimate the F0 of speech frames. In the second step, these frames are separated in low-pitch or high-pitch ones with FSFFE \cite{QUEIROZ_2022} technique. The separation leads to detection and adjustment of the F0 values according some typical errors  \cite{GAZOR_2020} that may occur in estimation, improving its accuracy. In sequence, the third stage consists in filtering the harmonic components of noisy speech with Gammachirp. The central frequencies and bandwidths of the filterbank are selectively defined to cover the most relevant regions for speech intelligibility, as stated in \cite{JENSEN_2016}. Finally, the filtered components are amplified by a gain factor to highlight the harmonic components of speech. This amplification mitigates the masking effects of background noise leading to intelligibility enhancement.


Several experiments are conducted to examine the effectiveness of the HDAG method. For this purpose, speech utterances collected from TIMIT \cite{TIMIT_1993} database are corrupted by six real acoustic noises, considering four SNR values: -10 dB, -5 dB, 0 dB and 5 dB. The proposed method and three baseline approaches are examined in terms of intelligibility enhancement. To this end, ESTOI \cite{JENSEN_2016} and Short-Time Approximated Speech Intelligibility Index (ASII$_\text{ST}$) \cite{TAAL_2015} are considered in the evaluation. Moreover, results for the Perceptual Evaluation of Speech Quality (PESQ) \cite{PESQ_2001} demonstrate that HDAG also achieve quality assessment. Objective results indicate that the proposal outperforms the competitive approaches in terms of speech intelligibility, and also quality scores. These results are corroborated by a subjective listening evaluation test. The main contributions of this work are:

\begin{itemize}

 \item Introduction of the HDAG method to improve the intelligibility and quality of acoustic noisy speech.

 \item Definition of the filterbank configuration using the third-octave bands and specific modulation frequencies, with higher resolution in regions most relevant to intelligibility.

 \item Adoption of the asymmetry coefficient from Gammachirp to adjust the filterbank to the noisy masked components of speech.
 
 \item Interesting intelligibility and quality assessment attained with adaptive gain factors defined according FSFFE separation.
 
\end{itemize}

The remaining of this paper is organized as follows. Section II describes the steps of the proposed HDAG method for intelligibility enhancement. An explanation of the competitive approaches SSFV, PACO (pitch-adaptive complex-valued Kalman filter) \cite{PEJMAN_2019}  and GTF$_\text{F0}$ is included in Section III. Section IV presents the evaluation experiments and results. Finally, Section V concludes this work.

\section{The HDAG Method}

The proposed method includes four main steps: harmonic detection, third-octave bands configuration, gammachirp filtering and output samples amplification by a gain factor. Finally, the overlap and add method is applied to achieve the reconstructed version of the target speech signal. Fig. \ref{block_diag} illustrates the block diagram of the HDAG method.

\subsection{F0 Estimation}

The fundamental frequency (F0) is estimated from noisy speech signal with HHT-Amp method \cite{HHT}. This F0 estimator ensures \cite{HHT}\cite{QUEIROZ_2022} interesting accuracy results from noisy speech signals. HHT-Amp is evaluated in a wide range of noisy scenarios outperforming four competing estimators in terms of accuracy. It applies the time-frequency EEMD (Ensemble Empirical Mode Decomposition) \cite{EMD_ORIG}\cite{EEMD_FLANDRIN} to decompose a voiced sample sequence $x_q(t)$ such that
\begin{equation}\label{eemd}
 x_q(t) = \sum_{k=1}^{K} \mathrm{IMF}_{k,q}(t) + r_q(t)
\end{equation}
where $\mathrm{IMF}_{k,q}(t)$ is the $k$-th mode of $x_q(t)$ and $r_q(t)$ is the last residual. Then, instantaneous amplitude functions are computed by
\begin{equation}
 a_{k,q}(t) = |Z_{k,q}(t)|, k = 1, \ldots, K,
\end{equation}
from the analytic signals defined as
\begin{equation}
 Z_{k,q}(t) = \mbox{IMF}_{k,q}(t) + j \, H\{\mbox{IMF}_{k,q}(t)\},
\end{equation}
where $H\{\mbox{IMF}_{k,q}(t)\}$ refers to the Hilbert transform of $\mbox{IMF}_{k,q}(t)$. The Autocorrelation Function is calculated as
\begin{equation}
 r_{k,q}(\tau) = \sum_t  a_k(t) \, a_k(t+\tau).
\end{equation}

For each decomposition mode $k$, let $\tau_0$ be the lowest $\tau$ value that correspond to an ACF peak, subject to $\tau_{min} \leq \tau_0 \leq \tau_{max}$. The frequency restriction is applied according to the range $[F_{min}, F_{max}]$ of possible F0 values. The $k$-th F0 candidate is defined as $\tau_0 / {f_s}$, where $f_s$ refers to the sampling rate. Finally, a decision criterion \cite{HHT} is applied to select the best pitch candidate $\hat T_0$. Finally, the estimated F0 is given by $ f_{est} = 1 / \hat T_0$.

\subsection{Harmonic Detection and Adjustment}

\begin{figure}[t!]
\centering
\includegraphics[width=\linewidth,keepaspectratio=true]{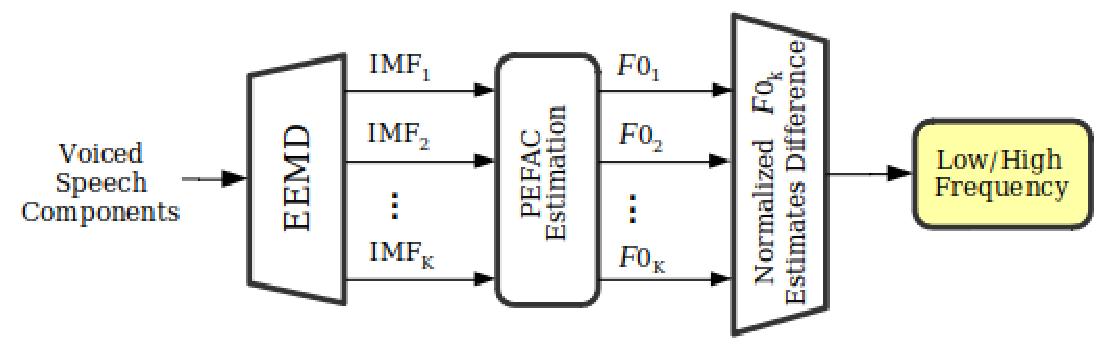}
\caption{Block diagram of the FSFFE technique for low/high pitch classification of speech frames.}
\label{fsffe_diag}
\end{figure}

Severe noise masking effects may impact the harmonic components of voiced speech leading to low accuracy F0 estimates. In order to detect and adjust the erroneous F0 values the FSFFE (Frequency Separation for Fundamental Frequency Estimation) \cite{QUEIROZ_2022} is applied to harmonic frames. This strategy separates the noisy speech frames into low-pitch or high-pitch ones. Possible errors in F0 estimates can be detected by comparing its values with the separation. Fig. \ref{fsffe_diag} illustrates the block diagram of the FSFFE method.

After the EEMD decomposition as in (\ref{eemd}), pitch estimation is performed in voiced frames of each IMF using PEFAC \cite{PEFAC_2011} algorithm. Let $\hat{F}0_{k,q}$ denote the pitch value estimated from frame $q$ of $\mathrm{IMF}_k \left ( t \right )$, the $\hat{F}0_{q}$ vector is composed as
\begin{equation}\label{pefac_eemd}
 \hat{F}0_{q} = \begin{bmatrix}
\hat{F}0_{1,q}, \hat{F}0_{2,q}, \cdots, \hat{F}0_{K,q}
\end{bmatrix}^T,
\end{equation}
to express the tendency that the frame is placed in a low/high pitch region. Only the first four IMFs ($K=4$) are considered in order to avoid the acoustic noise masking effect. The energy of these unwanted components are mostly concentrated at low frequencies ($K>6$) \cite{ZAO_2014}\cite{MEDINA_2021}\cite{CHATLANI_2012}.

A normalized distance is computed between IMFs for the successive frames to detect and overcome the differences in the estimated F0. Let $k$ and $k'$ denote IMF indexes, the distance is described as
\begin{equation}\label{norm_dist}
 \delta^q_{\hat{F}0}(k,k') = \left | \frac{\hat{F}0_{k,q} - \hat{F}0_{k',q}}{\hat{F}0_{k,q} + \hat{F}0_{k',q}} \right |.
\end{equation}
The $\delta^q_{\hat{F}0}(k,k')$ values are computed for different indexes of $k$ and $k'$ resulting in a 4x4 distance matrix $\delta^q_{\hat{F}0}$. The row components of the matrix are summed to obtain the variation property for the $k$-th IMF. The frequency region is defined as the mean value of PEFAC F0 estimates ($\bar{F}0_{q}$) between the two IMFs with the smallest variation scores. Finally, the low/high pitch separation is performed according the threshold $\gamma$ as
\begin{equation} \label{classif}
\begin{cases}
 \bar{F}0_{q} \leq \gamma, &  \text{low-frequency frame};\\
\bar{F}0_{q} > \gamma, &  \text{high-frequency frame}.
\end{cases}
\end{equation}
The threshold $\gamma$ is fixed in 200 Hz which is related to the average values between male (50-200 Hz) and female (120-350 Hz) speakers \cite{TITZE_1994}.

The F0 adjustment is conducted according the low/high pitch classification in (\ref{classif}). The F0 estimates are prone to doubling errors in low pitch frames. Hence, a low pitch frame that presents F0 value ($f_{est,q}$) ranging from [200-400]Hz is adjusted to $f_{adj,q} = 0.5f_{est,q}$. On the other hand, the high pitch frame is adjusted according possible halving and quartering \cite{GAZOR_2020} errors as follows:
\begin{equation}\label{criteriohigh}
 f_{adj,q} = \begin{cases}
4f_{est,q}, & 50 \leq f_{est,q} \leq 100\\ 
2f_{est,q}, & 100 <  f_{est,q} \leq 200
\end{cases}.
\end{equation}

\begin{figure}[t!]
\centering
\includegraphics[width=0.58\linewidth,keepaspectratio=true]{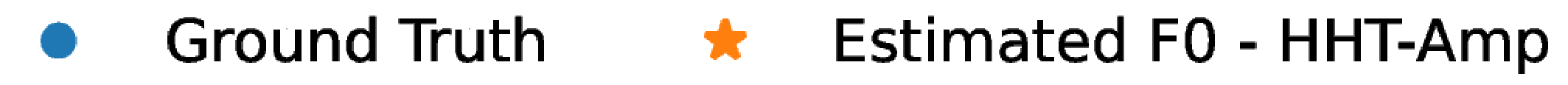}
\includegraphics[width=0.95\linewidth,keepaspectratio=true]{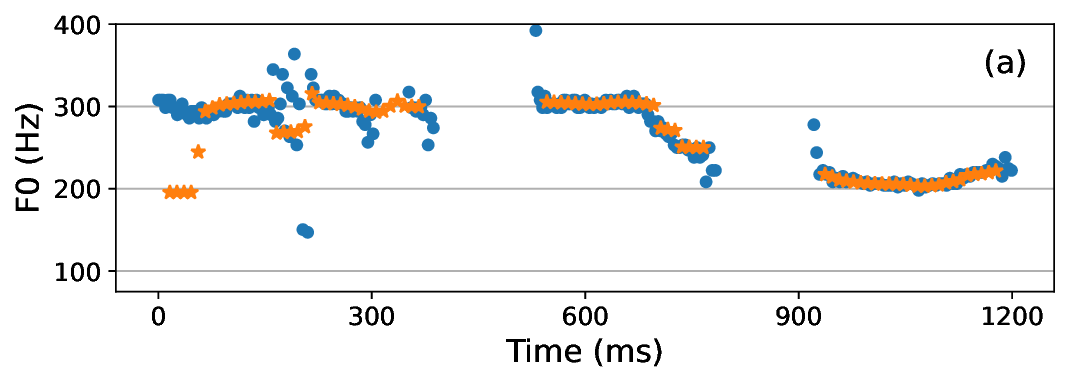}\vspace{-0.50cm}
\includegraphics[width=0.95\linewidth,keepaspectratio=true]{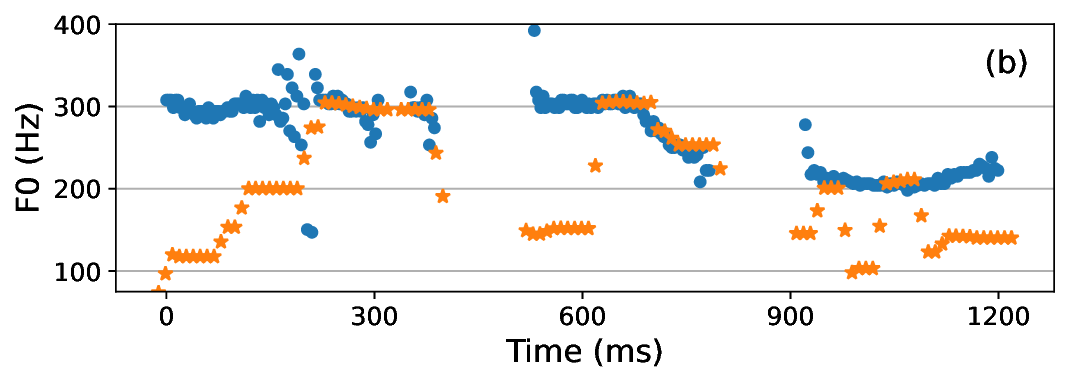}\vspace{-0.5cm}
\includegraphics[width=0.95\linewidth,keepaspectratio=true]{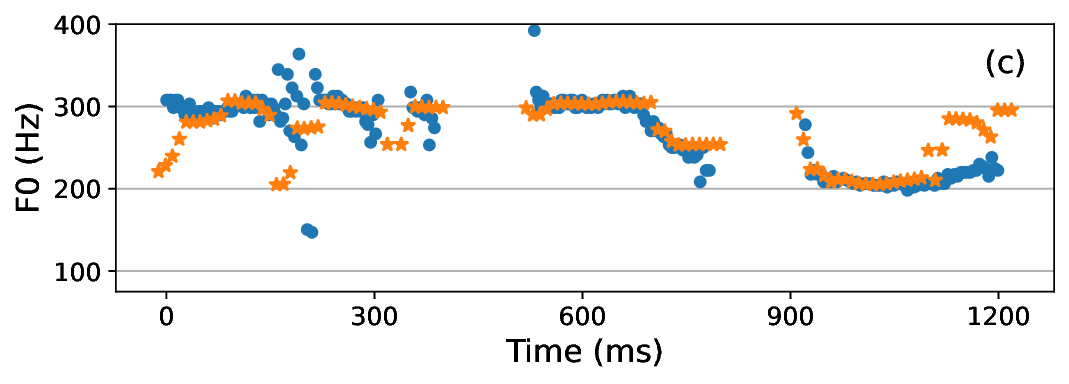}
\vspace{-0.1cm}
\caption{Ground Truth and F0 estimated with HHT-Amp technique for: (a) Clean Speech segment, (b) Noisy Signal with babble SNR=-5dB and (c) same Noisy segment with estimates improved by FSFFE.}
\label{f0_adjust}
\end{figure}

Fig. \ref{f0_adjust} illustrates the F0 adjustment procedure in frames of a 1200 ms speech signal. Fig. \ref{f0_adjust}(a) refers to F0 attained with HHT-Amp method for the clean speech. The estimated values match the ground truth in the high pitch region. Fig. \ref{f0_adjust}(b) presents the F0 estimates related to the noisy version of the same speech segment for the babble noise \cite{RSG10} with SNR = -5dB. Note that accuracy decreases significantly and halving errors appear in harmonic components, e.g., around 100 ms or 600 ms. These regions are adjusted with FSFFE as can be seen in Fig. \ref{f0_adjust}(b). Observe that the proposed adjust leads to accuracy improvement even in the severe noisy condition. The correction in harmonic detection is important specially in this case. Particularly, due to the fact that important components for speech intelligibility are placed in higher frequencies.

\subsection{Third-octave Bands Configuration}

Third-octave filter banks have been shown to loosely approximate the measured bands of the auditory filters \cite{IEC_1995}. Objective speech metrics consider the analysis of clean and noisy speech with third-octave subspaces. This is the case of ESTOI \cite{JENSEN_2016} intelligibility measure, that gives a prediction through the correlation of third-order spectrograms from the reference and processed signal.

This work proposes the definition of an auditory filtering based on the third-octave bands. The accurate harmonic detection $f_{adj,q}$ is adopted as center frequency of the first band of the filter bank ($k$ = 0). The center frequencies for the following $k$ bands are attained adaptively by
\begin{equation}\label{center_freq}
 f_c(k,q) = 2^{\frac{k}{3}}f_{adj,q}.
\end{equation}

The resulting set of filters in each frame $q$ provides better resolution in the frequencies near the harmonics of speech, which are the most important for the intelligibility \cite{JENSEN_2016}.

\subsection{Gammachirp Filtering}

\begin{table}[t!] \caption{\label{gammac_coef} ESTOI [$\times$10$^{-2}$] Scores for Different Asymmetry Coefficients.}
\centering
\renewcommand{\arraystretch}{1.25}
\setlength{\tabcolsep}{3.pt}
{
\begin{tabular}{crccccccccc}

\hline
&&\multicolumn{9}{c}{Gammachirp Coefficient -- $c$}\\\cline{3-11}

Noise&&2.0&1.5&1.0&0.5&0.0&-0.5&-1.0&-1.5&-2.0\\\hline

\multirow{5}{*}{Babble}&-10 dB&18.5&19.3&18.5&18.9&19.0&18.4&\bf19.4&18.4&19.1\\
&-5 dB&28.8&29.7&28.9&29.2&29.5&28.8&\bf29.8&28.8&29.6\\
&0 dB&41.1&42.0&41.3&41.6&41.8&41.1&\bf42.2&41.1&42.0\\
&5 dB&54.5&55.4&54.7&54.9&55.2&54.5&\bf55.6&54.4&55.5\\\cline{2-11}
\multicolumn{2}{r}{Average}&35.7&36.6&35.9&36.1&36.4&35.7&\bf36.8&35.7&36.5\\ \hline

\multirow{5}{*}{SSN}&-10 dB&20.5&21.3&20.6&20.9&21.0&20.4&\bf21.5&20.4&21.2\\
&-5 dB&30.2&31.1&30.4&30.7&30.8&30.3&\bf31.3&30.1&31.1\\
&0 dB&41.6&42.4&41.8&42.0&42.2&41.6&\bf42.6&41.5&42.5\\
&5 dB&54.2&55.0&54.4&54.6&54.8&54.2&\bf55.3&54.1&55.2\\\cline{2-11}
\multicolumn{2}{r}{Average}&36.6&37.4&36.8&37.0&37.2&36.6&\bf37.7&36.5&37.5\\ \hline

\end{tabular}
}
\end{table}

In this step, a set of $L$ Gammachirp filters \cite{PATTERSON_1997} $\left\{ h_k(t), k=1 \ldots, L \right\}$ are applied to successively filter the input sample sequence $x_q(t)$. Each filter $h_k(t)$ is implemented to the noisy signal considering frames of 32 ms, order $n = 4$, center frequencies given by (\ref{center_freq}). In order to align the impulse response functions, phase compensation is applied to all filters, which correspond to the non-causal filters
\begin{equation}
h_k(t) = a (t+t_c)^{n-1} \cos(2 \pi f_c t + c  \text{ln}t) e^{-2 \pi b (t+t_c)}\, , \, t \geq -t_c\, ,
\label{eq:gamma_c}
\end{equation}
where $c$ is the gammachirp coefficient of the filter and $t_c = \frac{n-1}{2 \pi b}$, which ensures that peaks of all filters occur at $t = 0$.

The bandwidth $b$ is defined here according the frequencies of modulation transfer function considered in \cite{PLOMP_1994}\cite{PLOMP_1994_2}\cite{THEUNISSEN_2009}. The results presented in \cite{THEUNISSEN_2009} demonstrated that the frequency range relevant for intelligibility of male speech sentences ranges from [1--12.5] Hz. Nevertheless, female sentences presented a larger range, with noticeable relevance for frequencies $\leq$ 20 Hz. Therefore, this work proposes an harmonic-adaptive bandwidth, given by $b = 0.15 f_{adj,q}$.

Let $x_q^0(t) = x_q(t)$, the filtered signals $y_q^k(t), k = 1, \ldots, L$, are recursively computed by
\begin{equation}
\left\{\begin{array}{l}
y_q^k(t) = x_q^{k-1}(t) * h_k(t) \vspace{0.2cm}\\
x_q^k(t) = x_q^{k-1}(t) - y_q^k(t)
\end{array} \right. ,
\quad k = 1, \ldots, L\, .
\label{eq:filters}
\end{equation}
The residual signal is defined as $r_q(t) = x_q^L(t)$ to guarantee the completeness of the input sequence, i.e.,
\begin{equation}
 x_q(t) = \sum_{k=1}^L y_q^k(t) + r_q(t).
\end{equation}

Table \ref{gammac_coef} presents the ESTOI scores for different asymmetry coefficients $c$ of the Gammachirp filter. The intelligibility is predicted for a training subset of 48 speech signals of TIMIT \cite{TIMIT_1993} defined in \cite{GONZALEZ_2014}. The ESTOI scores with different values of $c$ is computed for Babble \cite{RSG10} and SSN \cite{DEMAND_2013} noisy scenarios. Note that the coefficient $c$ = -1 achieves the highest intelligibility rates for all the noisy conditions. This can be justified by the fact that acoustic noises might shift the harmonic detection. Therefore, the asymmetry of gammachirp has the role of fine-tuning in these harmonic components.

\begin{figure}[t!]
\begin{center}
 \includegraphics[width=0.5\linewidth,keepaspectratio=true, clip=true,trim=0pt 0pt 0pt 0pt]{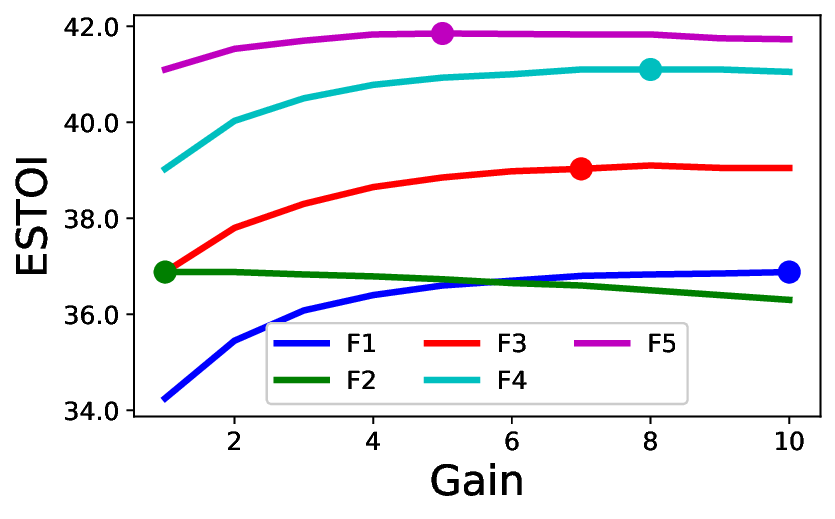}
\hspace{-0.3cm}
 \includegraphics[width=0.5\linewidth,keepaspectratio=true, clip=true,trim=0pt 0pt 0pt 0pt]{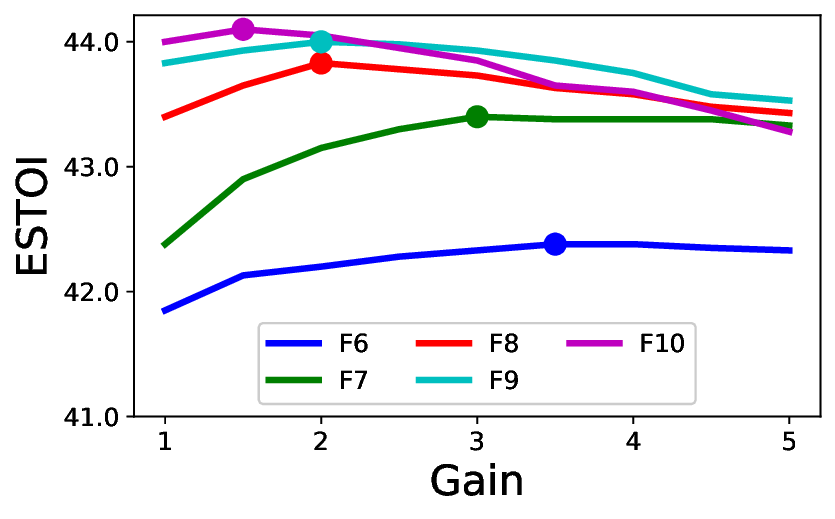}
(a)
\\
 \includegraphics[width=0.5\linewidth,keepaspectratio=true, clip=true,trim=0pt 0pt 0pt 0pt]{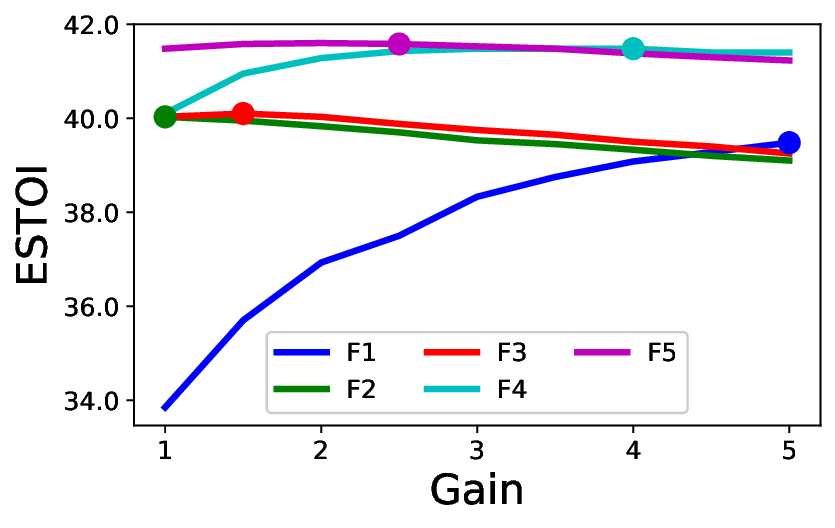}
\hspace{-0.3cm}
 \includegraphics[width=0.5\linewidth,keepaspectratio=true, clip=true,trim=0pt 0pt 0pt 0pt]{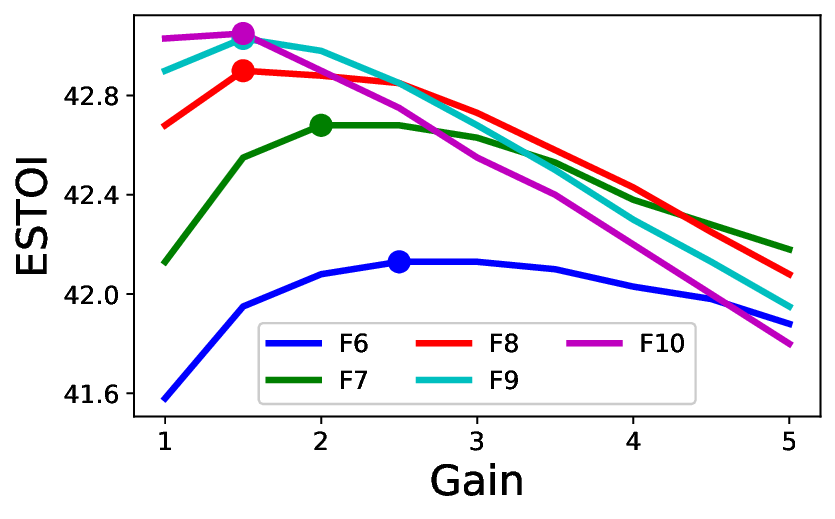}
 \vspace{-0.05cm}
(b)
\caption{ESTOI curves of (a) low pitch and (b) high pitch frames averaged for SNR values: -10dB, -5dB, 0dB and 5dB of Babble noise according the gain factor $G_k$ for each gammachirp filter.}
\label{gain_factor}
\end{center}
\end{figure}

\subsection{Frames Reconstruction with a Gain Factor}

After the Gammachirp filtering, the amplitude of the output samples $y_q^k(t), k = 1, \ldots, L$, are amplified by a gain factor $G_k \geq 1$. The idea is to emphasize the presence of the harmonic features of speech, which will lead to speech intelligibility improvement, without introducing any noticeable distortion to the speech signal. The reconstruction of the voiced frame $q \in S_v$ leads to the sample sequence

\begin{equation}
\hat x_q(t) = \left[ \sum_{k=1}^L G_k \, y_q^k(t) \right] + r_q(t)\, .
\label{eq:frame_rec}
\end{equation}

The reconstructed voiced frames in $S_v$ and all the remaining frames in $S_u$ are joined together keeping the original frames indices. Thus, all frames are overlap and added to reconstruct the modified version $\hat x(t)$ of the target speech signal. The completeness and continuity of $\hat x(t)$ is guaranteed by the adoption of the Hanning window that multiply all frames before the overlap and add method. This means that the reconstructed signal $\hat x(t)$ and the original signal $x(t)$ would be exactly the same if each frame is reconstructed considering $G_k = 1$ for every $k \in \left\{1, \ldots, L\right\}$.

The set of gains $G_k$ are empirically determined in each filter using the same training subset of 48 speech signals attained from TIMIT database. Fig. \ref{gain_factor} illustrates the ESTOI curves for noisy speech signal with Babble and averaged to four SNR values. The configuration starts from the first filter (F1), and the gain is incremented until ESTOI reaches its maximum value (highlighted point). This gain is fixed, and the process is repeated for the subsequent filters. Observe that two different sets of gain are presented: one for low pitch (Fig. \ref{gain_factor}(a)) and other for high pitch frames (Fig. \ref{gain_factor}(b)). Therefore, the $G_k$ values for $L$ = 10 filters that lead to the highest intelligibility ESTOI scores are defined as

\begin{equation} \label{gain_sets}
\hspace{-0.3cm} G_k = 
\begin{cases}
\{14,1,4,8,4,3.5,3,2,2,1.5\}, & \hspace{-0.15cm} \text{low-pitch};\\
\{14,1,1,4.5,2,3.5,2.5,2,1.5,1.5\}, & \hspace{-0.15cm} \text{high-pitch}.
\end{cases}
\end{equation}

The proposed HDAG method is summarized in Algorithm \ref{alg:alg1}. This algorithm is tailored to the harmonic detection scheme considered in this paper. However, Algorithm 1 can be also used with any other F0 estimation technique.

\begin{algorithm}[t!]
\caption{Intelligibility Enhancement Scheme HDAG.}\label{alg:alg1}
\begin{algorithmic}
\STATE\textbf{for} $q$ \textbf{do}
\STATE\setlength{\leftskip}{0.5cm}Input: $x_q(t)$
\STATE\textbf{Harmonic Detection}
\STATE$f_{est,q} \gets$ F0 estimation with HHT-Amp as in Section II-A.
\STATE$\hat{F}0_q \gets$ PEFAC (\ref{pefac_eemd}) for $K$=4 decomposed modes of (\ref{eemd}).
\STATE$\delta^q_{\hat{F}0} \gets$ normalized distance matrix using (\ref{norm_dist})
\STATE low/high pitch classification (\ref{classif}) according $\bar{F}0_{q}$.

\STATE\textbf{Gammachirp Filtering}
\STATE\textbf{for} $k$ \textbf{do}
\STATE\setlength{\leftskip}{1cm}$h_k(t) \gets$ impulse response of non-causal filters (\ref{eq:gamma_c})
\STATE$y_q^k(t) = x_q^{k-1}(t) * h_k(t)$
\STATE$x_q^k(t) = x_q^{k-1}(t) - y_q^k(t)$
\STATE\setlength{\leftskip}{0.5cm}\textbf{end for}
\STATE$r_q(t) = x_q^L(t) \gets$ residual components
\STATE$\hat x_q(t) \gets$ voiced frames reconstruction as in (\ref{eq:frame_rec}) and $G_k$ from (\ref{gain_sets}).
\STATE$\hat x(t) \gets$ overlap and add technique.

\STATE\setlength{\leftskip}{0cm}\textbf{end for}
\STATE\textbf{return} $\hat x(t)$
\end{algorithmic}
\label{alg1}
\end{algorithm}

\section{Harmonic-based Comparative Methods}

This Section briefly describes the baseline methods SSFV, PACO and GTF$_\text{F0}$. They also consider the harmonic components of noisy speech to attain intelligibility and quality improvement.

\subsection{SSFV}

The main idea of this solution consists on transforming the original signal, adopting a Lombard effect strategy \cite{LOMBARD_1911} \cite{JUNQUA_1993}. In this effect the central frequencies of the formants are shifted (Formant Shifting). It moves away the energy from these frequencies from the region of spectral action of the noise. The formant shifting process is described in \cite{GAEL_2016} and optimized to operate in environments with the presence of Car noise (composed by radio, message alert and telephone). Initially, LPC (Linear Prediction Coding) is used to estimate the poles and formant frequencies of the voiced speech signal. In the LPC model, a 25ms frame of the signal $s(n,m)$ can be represented by linear predictions of order $p$ \cite{RABINER_1978}, that is
 \begin{equation}
  s(n,m)=\sum_{j=1}^{p}a_js(n-j,m)+e(n,m),
 \end{equation}
where $a_j$ are the linear prediction coefficients, $e(n,m)$ indicates the residual error and $p=12$. The variables $n$ and $m$ represent the signal sample and time frame indices, respectively. The LP filter A($z$) is obtained from the coefficients $a_j$, so that
 \begin{equation}
  A(z)=1+\sum_{j=1}^{p}a_jz^j.
 \end{equation}
The poles \textbf{P} are obtained by the roots of the LP coefficients, and the formant frequencies \textbf{F} are defined as the estimated pole angles.

The formants obtained are shifted according to a function $\delta(F)$ \cite{GAEL_2017} determined according to the characteristics of the acoustic noise. The displacement of formants is carried out according to the criterion
 \begin{equation}
    \hat{F}(f) = \left\{\begin{matrix}
    F(f)+\delta(f), \quad f_1<f<f_3 \\
    F(f), \quad \quad \quad \quad \text{otherwise}.
    \end{matrix}\right.
 \end{equation}
where $f_1$ and $f_3$ are the first and third formants, respectively. Finally, the resulting set of formants $\hat{\textbf{F}}$ is obtained from these modifications.

\vspace{-0.2cm}
\subsection{PACO}

The pitch-adaptive complex-valued Kalman filter (PACO) \cite{PEJMAN_2019} is also adopted as a competitive technique for the proposed HDAG method. It applies the harmonic signal modeling for estimating the complex-valued speech AR parameters required for the Kalman filter. To this end, fundamental frequency estimation $f$ for each 32ms signal frame $y(n,l)$ is performed and phase progression $\hat{\psi}(l)$ is recursively estimated for the harmonic $h$ according to
\begin{equation}
 \psi_h(l) = \psi_h(l-1) + \frac{\pi L}{f_s}(f_h(l) + f_h(l-1)).
\end{equation}

Successive speech DFT bins of $y(n,l)$ is computed by incorporating the harmonic phase progression into a state-transition model. The AR coefficients $\hat{\bf a}(l)$ are defined from the DFT bins \cite{PEJMAN_2019}, which are the input for the Kalman filter gain $G_K$ and obtain an estimation of $\hat{\bf X}(k,l)$ such as
\begin{equation}
 \hat{X}(k,l) = G_k(k,l)(Y(k,l) - \hat{X}_{prop}(k,l))
\end{equation}
where $\hat{X}_{prop}$ is the state propagation estimate for the k-$th$ bin. Finally, inverse DFT is applied and the processed speech signal is reconstructed performing overlap and add.

\begin{table*}[t!] \caption{\label{estoi_pesq} Intelligibility and Quality results with the Proposed HDAG and Competitive methods.}

\renewcommand{\arraystretch}{1.2}
\setlength{\tabcolsep}{9.pt}
\begin{center}
{
\begin{tabular}{crccccccccccc}

\hline
&&\multicolumn{5}{c}{ESTOI}&&\multicolumn{5}{c}{PESQ}\\
\cline{3-7}\cline{9-13}
Noise&\multicolumn{1}{c}{SNR}&UNP&SSFV&PACO&GTF$_\text{F0}$&HDAG&&UNP&SSFV&PACO&GTF$_\text{F0}$&HDAG\\\hline

\multirow{5}{*}{Babble} &-10 dB&0.18&0.17&0.17&0.24&\bf0.28&&0.56&0.99&1.25&1.77&\textbf{2.06}\\
&-5 dB&0.29&0.29&0.30&0.37&\bf0.40&&1.52&1.54&1.94&2.30&\textbf{2.50}\\
&0 dB&0.41&0.42&0.43&0.50&\bf0.53&&1.90&1.92&2.45&2.72&\textbf{2.86}\\
&5 dB&0.55&0.55&0.58&0.64&\bf0.66&&2.35&2.36&2.94&3.12&\textbf{3.22}\\
&Average&0.36&0.36&0.37&0.44&\bf0.47&&1.58&1.70&2.14&2.48&\textbf{2.66}\\ \hline

\multirow{5}{*}{Cafeteria} &-10 dB&0.20&0.19&0.19&0.27&\bf0.31&&1.00&1.30&1.50&1.97&\textbf{2.20}\\
&-5 dB&0.31&0.31&0.31&0.40&\bf0.43&&1.63&1.67&2.01&2.48&\textbf{2.63}\\
&0 dB&0.44&0.44&0.45&0.54&\bf0.56&&2.07&2.09&2.52&2.90&\textbf{3.02}\\
&5 dB&0.58&0.58&0.61&0.67&\bf0.69&&2.50&2.51&2.97&3.29&\textbf{3.37}\\
&Average&0.38&0.38&0.39&0.47&\bf0.50&&1.80&1.89&2.25&2.66&\textbf{2.81}\\ \hline

\multirow{5}{*}{Traffic} &-10 dB&0.38&0.38&0.44&0.44&\bf0.47&&1.59&1.58&\bf2.82&2.34&2.51\\
&-5 dB&0.51&0.50&0.56&0.58&\bf0.60&&2.04&2.04&\bf3.27&2.73&2.88\\
&0 dB&0.63&0.63&0.68&0.70&\bf0.71&&2.55&2.55&\bf3.62&3.12&3.26\\
&5 dB&0.74&0.74&0.79&0.79&\bf0.80&&3.06&3.06&\bf3.86&3.52&3.62\\
&Average&0.48&0.48&0.52&0.56&\bf0.58&&2.31&2.31&\bf3.39&2.93&3.07\\ \hline

\multirow{5}{*}{Train} &-10 dB&0.32&0.30&0.36&0.38&\bf0.42&&1.33&1.38&1.92&2.09&\textbf{2.29}\\
&-5 dB&0.43&0.43&0.47&0.51&\bf0.54&&1.82&1.83&2.55&2.61&\textbf{2.75}\\
&0 dB&0.55&0.54&0.58&0.63&\bf0.65&&2.33&2.34&3.03&3.06&\textbf{3.18}\\
&5 dB&0.65&0.65&0.69&0.73&\bf0.75&&2.78&2.79&3.37&3.42&\textbf{3.54}\\
&Average&0.49&0.48&0.53&0.56&\bf0.59&&2.06&2.08&2.72&2.79&\textbf{2.94}\\ \hline

\multirow{5}{*}{Helicopter} &-10 dB&0.30&0.30&0.33&0.39&\bf0.43&&1.55&1.59&2.21&2.34&\textbf{2.54}\\
&-5 dB&0.41&0.41&0.45&0.52&\bf0.54&&1.89&1.91&2.71&2.74&\textbf{2.87}\\
&0 dB&0.53&0.53&0.59&0.64&\bf0.66&&2.33&2.34&3.17&3.15&\textbf{3.26}\\
&5 dB&0.66&0.65&0.72&0.75&\bf0.76&&2.76&2.76&3.53&3.51&\textbf{3.60}\\
&Average&0.47&0.47&0.52&0.58&\bf0.60&&2.13&2.15&2.91&2.93&\textbf{3.06}\\\hline

\multirow{5}{*}{SSN} &-10 dB&0.17&0.16&0.20&0.24&\bf0.29&&1.22&1.41&1.95&1.88&\textbf{2.17}\\
&-5 dB&0.28&0.28&0.32&0.37&\bf0.41&&1.45&1.47&2.41&2.25&\textbf{2.41}\\
&0 dB&0.41&0.41&0.45&0.51&\bf0.54&&1.84&1.85&\bf2.89&2.68&2.80\\
&5 dB&0.54&0.54&0.59&0.64&\bf0.66&&2.32&2.33&\bf3.29&3.11&3.20\\
&Average&0.35&0.35&0.39&0.44&\bf0.47&&1.70&1.77&2.63&2.48&\textbf{2.65}\\ \hline
\multicolumn{2}{c}{Overall}&0.44&0.43&0.47&0.52&\bf0.54&&1.93&1.98&2.67&2.71&\textbf{2.86}\\\hline
\end{tabular}
}
\end{center}
\end{table*}

\vspace{-0.3cm}
\subsection{GTF$_\text{F0}$}

In the GTF$_\text{F0}$ \cite{QUEIROZ_2021} method, a set of $L$ Gammatone filters $\left\{ h_k(t), k=1 \ldots, L \right\}$ are applied to successively filter the input sample sequence $x_q(t)$. Each filter $h_k(t)$ is implemented\footnote{Code available at http://staffwww.dcs.shef.ac.uk/people/N.Ma/} in frames of 32 ms considering order $n = 4$, center frequency
\begin{equation}
 f_c = k F0
\end{equation}
and bandwidth $b = 0.25 F0$. The time-domain impulse response function described in (\ref{eq:gamma_c}) is applied for GTF$_\text{F0}$ without the asymmetry coefficient.  Thus, it can be considered a specific case of Gammachirp filterbank, in which $c$ = 0.

After the Gammatone filtering, the amplitude of the output samples $y_q^k(t), k = 1, \ldots, L$, are amplified by the following a gain factor $G_k \geq 1$. The integer multiples of F0 are amplified as in \cite{QUEIROZ_2021} with the following linear gains: $G_1$ = $G_2$ = 5.0, $G_3$ = 4.0 and $G_4$ = 2.5.

\section{Results and Discussion}

This section presents objective results for intelligibility and quality of acoustic signals processed by HDAG method in comparison to SSFV, PACO and GTF$_\text{F0}$ baseline techniques. ESTOI \cite{JENSEN_2016} and ASII$_\text{ST}$ \cite{TAAL_2015} are considered to evaluate the speech intelligibility improvement and PESQ \cite{PESQ_2001} compares the quality assessment of competitive methods. Following, results for a perceptual test is presented, in order to corroborate the objective evaluation.

The experimental scenario consider a subset\footnote{Available at: http://www.ee.ic.ac.uk/hp/staff/dmb/data/TIMITfxv.zip.} of TIMIT \cite{TIMIT_1993} database to evaluate the competitive methods. The set considered is composed by 128 speech signals spoken by 8 male and 8 female speakers, sampled at 16 KHz and with 3 s average duration. The F0 reference values and voiced/unvoiced information for the training and test datasets are obtained from \cite{GONZALEZ_2014}. Six noises are used to corrupt the speech utterances: acoustic Babble and Traffic attained from RSG-10 \cite{RSG10}, Cafeteria, Train and Helicopter from Freesound.org\footnote{[Online]. Available: https://freesound.org.}, and Speech Shaped Noise (SSN) from DEMAND \cite{DEMAND_2013} database. Experiments are conducted considering noisy signals with four SNR values (-10 dB, -5 dB, 0 dB and 5 dB). In this study, it is assumed that the FSFFE separation into high-pitch and low-pitch speech frames is considered perfect and generates no errors into the whole system.

\begin{figure*}

\begin{center}
 \includegraphics[width=0.33\linewidth,keepaspectratio=true]{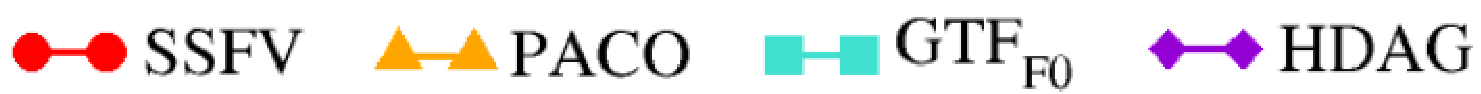}

\subfigure[]{
 \includegraphics[width=0.326\linewidth,keepaspectratio=true]{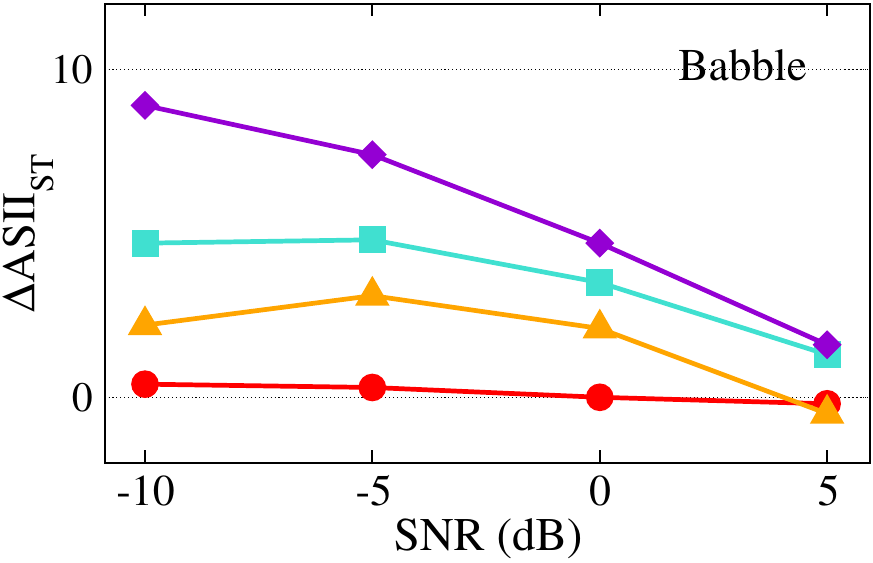}
}
\hspace{-0.4cm}
\subfigure[]{
 \includegraphics[width=0.326\linewidth,keepaspectratio=true]{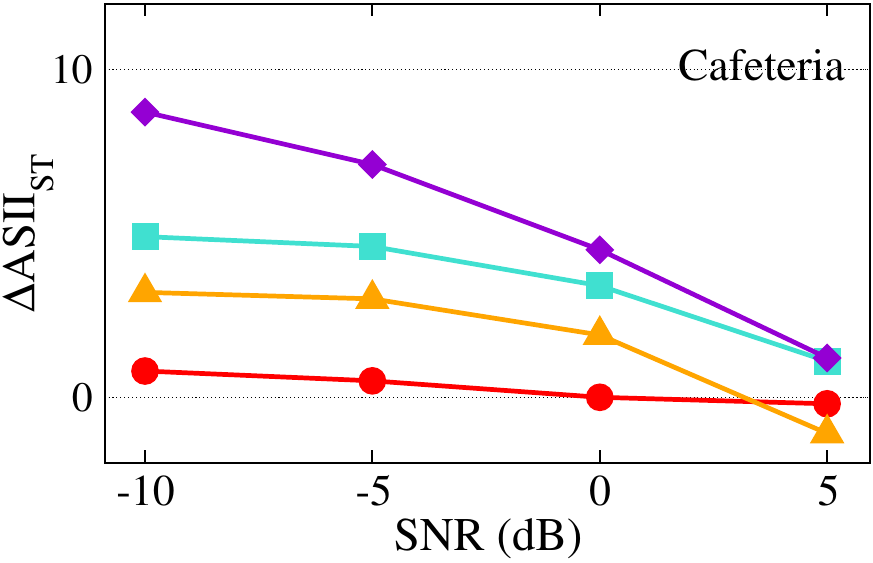}
}
\hspace{-0.4cm}
\subfigure[]{
 \includegraphics[width=0.326\linewidth,keepaspectratio=true]{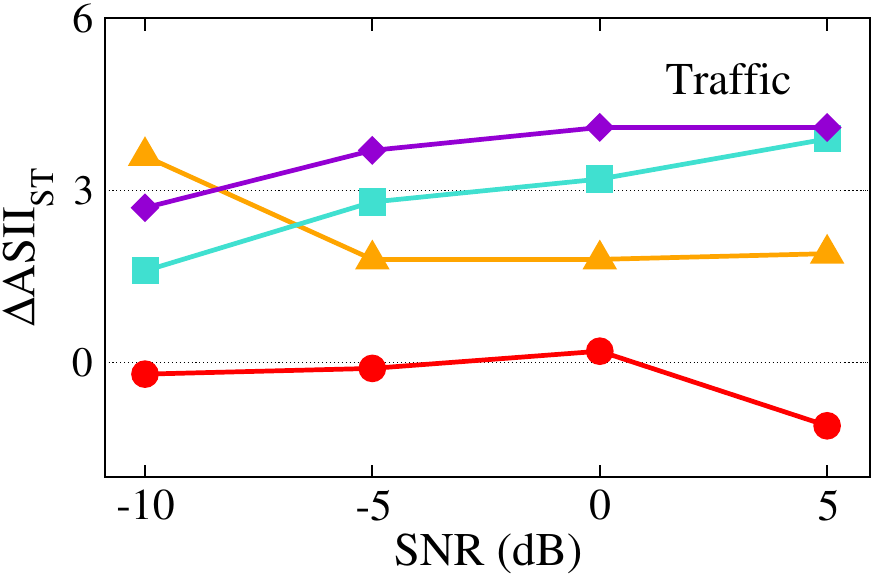}
}
\hspace{-0.4cm}
\subfigure[]{
 \includegraphics[width=0.326\linewidth,keepaspectratio=true]{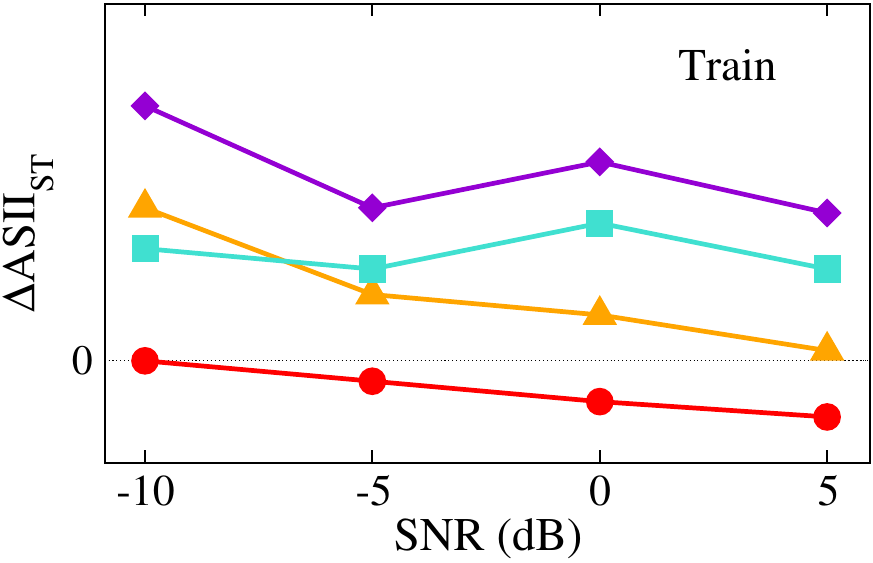}
}
\hspace{-0.4cm}
\subfigure[]{
 \includegraphics[width=0.326\linewidth,keepaspectratio=true]{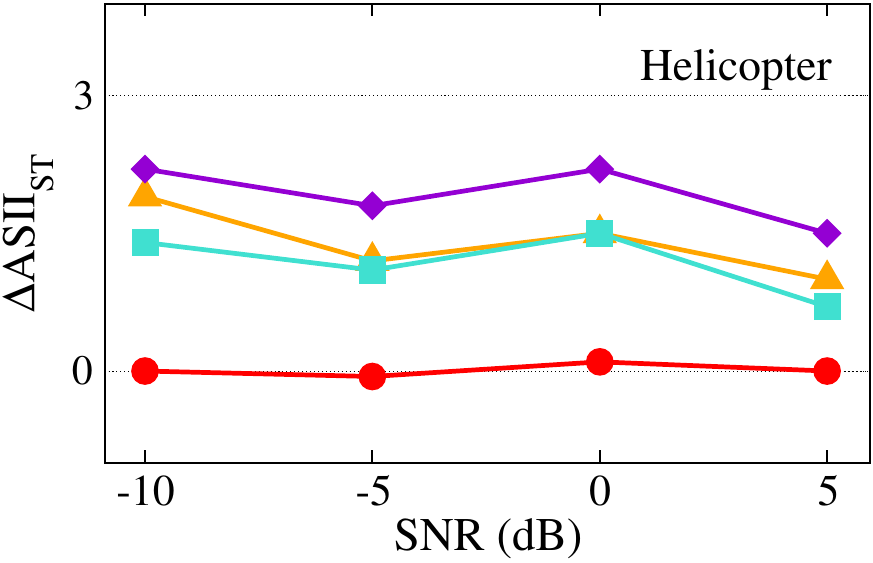}
}
\hspace{-0.4cm}
\subfigure[]{
 \includegraphics[width=0.326\linewidth,keepaspectratio=true]{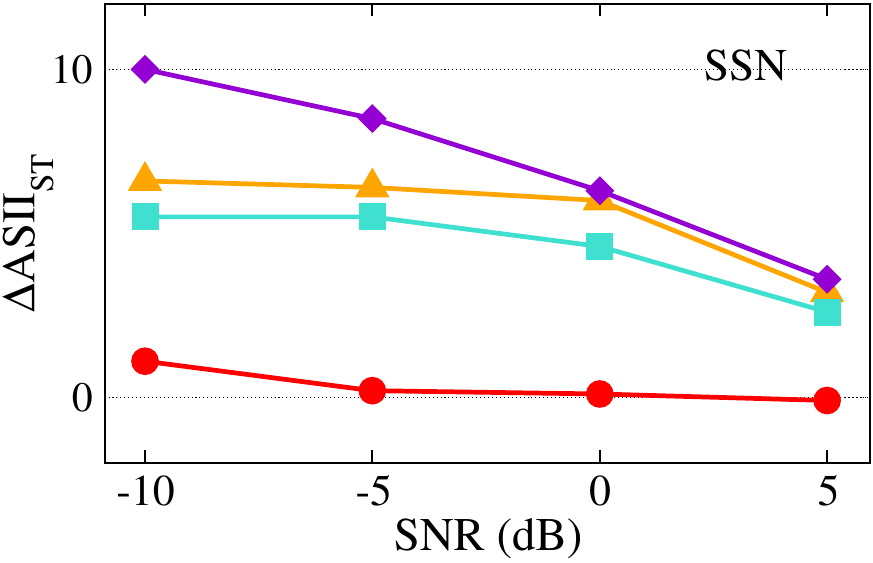}
}
\caption{$\Delta$ASII$_\text{ST}$ intelligibility enhancement [$\times$10$^{-2}$] averaged for speech signals corrupted by noises: (a) Babble, (b) Cafeteria, (c) Traffic, (d) Train, (e) Helicopter and (f) SSN.}
\label{asii}
\end{center}
 
\end{figure*}

\subsection{Intelligibility and Quality Objective Evaluation}

Table \ref{estoi_pesq} shows the intelligibility and quality objective results with ESTOI and PESQ measures, respectively. Note that Babble and SSN noises present the most challenging scenarios among those evaluated in terms of intelligibility. For instance, the ESTOI averaged for the SNR values of UNP speech signals are 0.36 and 0.35 for the respective noises. Moreover, observe that HDAG method achieves the best results for all the 24 noise conditions even in the most challenging scenarios with negative SNR values. The scores of HDAG are particularly interesting for the non-stationary noises, i.e., Babble and Cafeteria. For these noise sources the ESTOI attained are considerably higher than all the competing solutions for all SNR values. The highest ESTOI accomplished by HDAG is 13 p.p can be observed for Helicopter noise with SNR = -10 dB. According the overall average, the proposed solution outperforms the competitive approaches with ESTOI of 0.54, against 0.52, 0.47 and 0.43 for GTF$_\text{F0}$, PACO and SSFV, respectively.

The PESQ score is here computed from 30\% of the most relevant harmonic frames of noisy speech. These frames are selected from those with the lowest signal-to-noise ratio values. Note that HDAG outperforms the competing approaches for most of the noisy speech conditions in terms of quality assessment. The proposed solution achieves the highest PESQ, except for Traffic and SSN (0 dB and 5 dB) noises. For instance, in Helicopter with SNR = -10 dB the PESQ score attained by HDAG is 1.02 higher than UNP followed by increments of 0.79, 0.66 and 0.04 presented by GTF$_\text{F0}$, PACO and SSFV, respectively. In summary, the overall PESQ obtained with HDAG is 2.86, against 2.71 for the competing approach GTF$_\text{F0}$. Therefore, these results indicate that the proposed solution also provides quality assessment.

\begin{table}[t!] \caption{\label{asii_unp} ASII$_\text{ST}$ [$\times$10$^{-2}$] Scores for UNP noisy speech.}
\renewcommand{\arraystretch}{1.4}

\begin{center}
{
\begin{tabular}{rcccccc}

\hline
SNR&Babble&Cafeteria&Traffic&Train&Helicopter&SSN\\\hline

-10 dB&23.1&24.3&35.8&34.6&34.1&19.3\\
-5 dB&26.6&27.9&39.2&39.9&40.2&23.7\\
0 dB&33.0&34.3&43.7&47.2&43.5&30.0\\
5 dB&40.9&42.3&47.5&54.9&51.0&37.9\\\hline
Average&30.9&32.2&41.6&44.2&42.2&27.7\\\hline
\end{tabular}
}
\end{center}
\end{table}

Table \ref{asii_unp} presents the average ASII$_\text{ST}$ results for the unprocessed (UNP) noisy speech signals. Here the SSN and Babble noises attained the lowest scores for SNR value of -10 dB, with ASII$_\text{ST}$ of 19.3 and 23.1, respectively. The ASII$_\text{ST}$ values incremented by each competitive method ($\Delta$ASII$_\text{ST}$) are depicted in Fig. \ref{asii} for the six acoustic noises. Observe that the proposed solution accomplishes the highest scores for most conditions, except for Traffic (SNR = -10 dB). The best $\Delta$ASII$_\text{ST}$ (10.1$\times$10$^{-2}$) is achieved by the challenging SSN noise in -10 dB. As can be seen in ESTOI, the SSFV approach do not present noticeable ASII$_\text{ST}$ increment. Moreover, for the non-stationary Cafeteria noise the proposed solution attains average intelligibility enhancement of 5.4$\times$10$^{-2}$, compared with 3.5$\times$10$^{-2}$, 1.8$\times$10$^{-2}$ and 0.3$\times$10$^{-2}$ for baselines GTF$_\text{F0}$, PACO and SSFV. Therefore, these results reinforce the robustness of the proposed method against several noisy masking effects.

\begin{figure}

\begin{center}

\subfigure[]{
 \includegraphics[width=\linewidth,keepaspectratio=true, clip=true,trim=30pt 0pt 50pt 0pt]{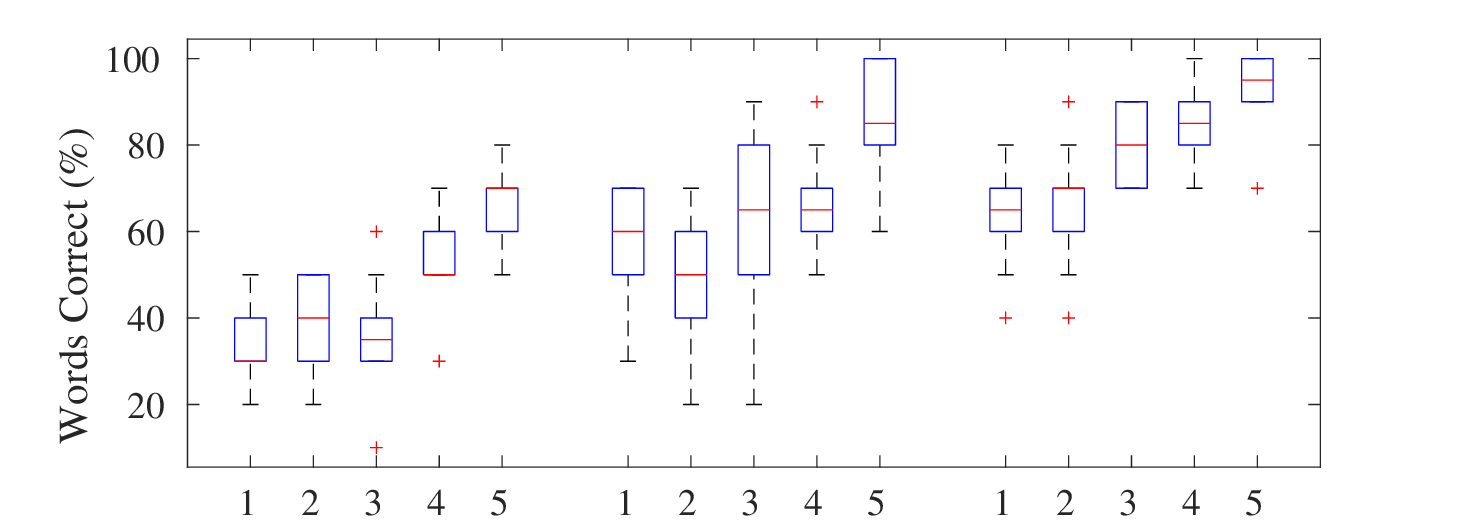}
}
\hspace{-0.01cm}
\subfigure[]{
 \includegraphics[width=\linewidth,keepaspectratio=true, clip=true,trim=30pt 0pt 50pt 0pt]{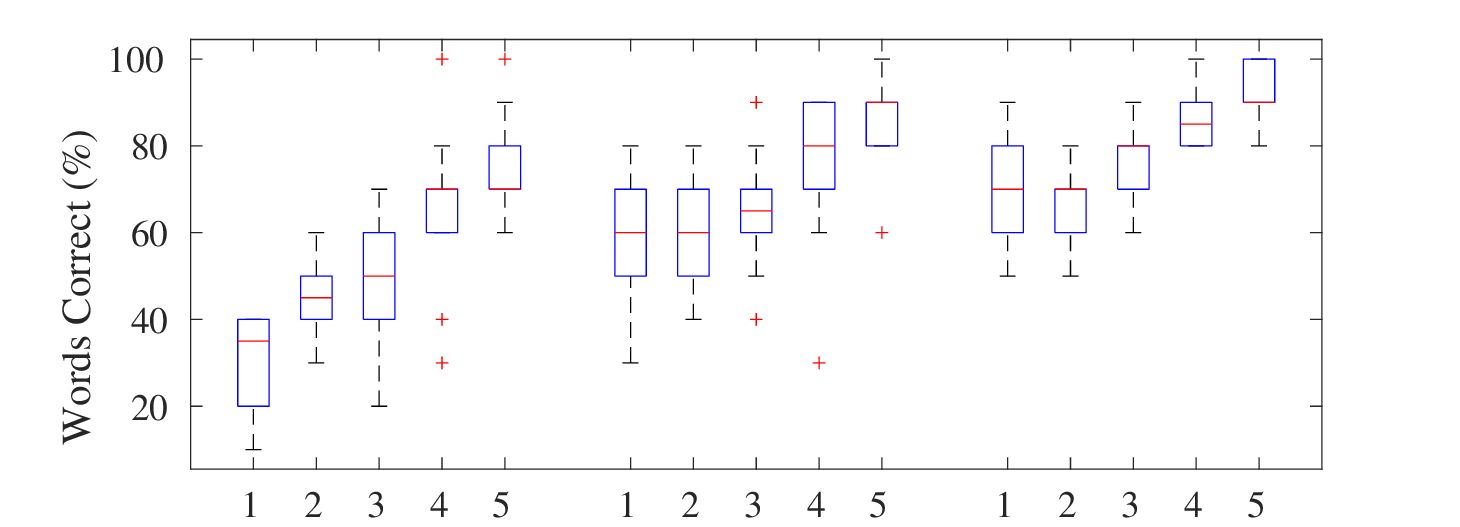}
}
\hspace{-0.01cm}
\subfigure[]{
 \includegraphics[width=\linewidth,keepaspectratio=true, clip=true,trim=30pt 0pt 50pt 0pt]{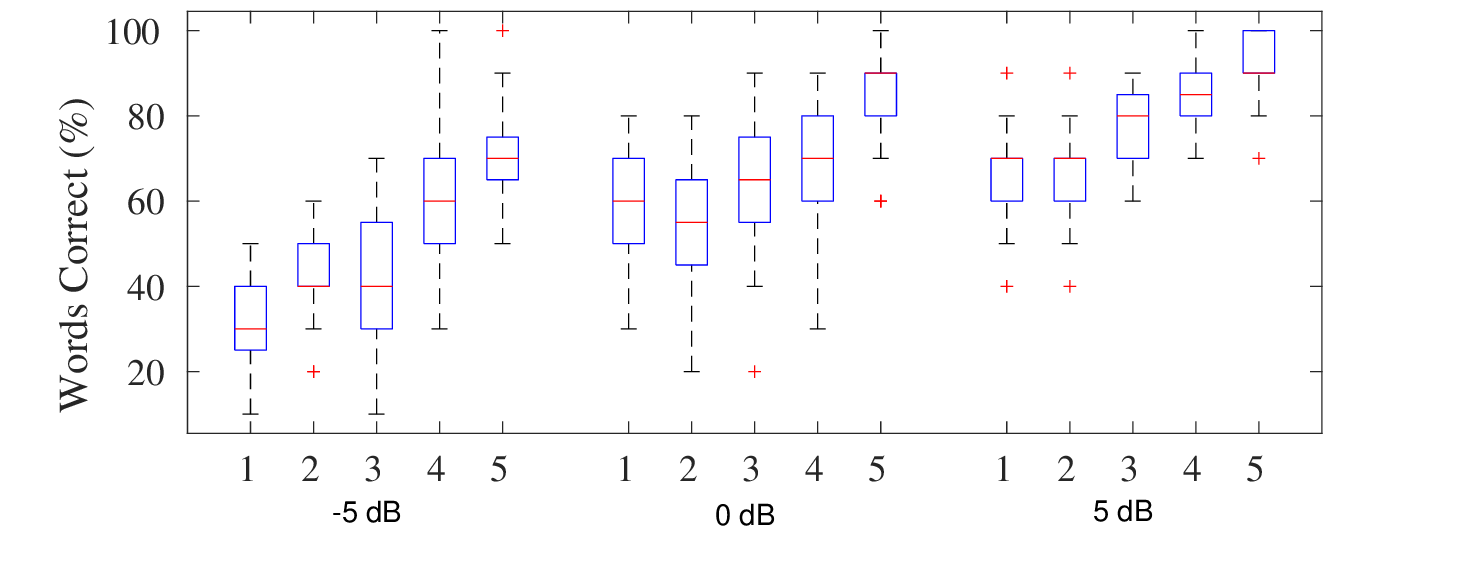}
}
\caption{Perceptual intelligibility evaluation with SSN additive acoustic noise for (a) male, (b) female volunteers and (c) overall scores. Each case denotes: 1-UNP, 2-SSFV, 3-PACO, 4-GTF$_\text{F0}$ and 5-HDAG}
\label{percep_test}
\end{center}
 
\end{figure}

\subsection{Perceptual Intelligibility Evaluation}

A subjective listening test \cite{GHIMIRE_2012} is conducted considering a scenario of phonetic balanced words\footnote{The complete test database is available at lasp.ime.eb.br.}. Ten native male and ten female Brazilian volunteers perform the test, which ages range from 19 to 57 years with an average of 32. The SSN noise is adopted with SNRs of -5 dB, 0 dB and 5 dB. Ten words are applied for each of the 15 test conditions, i.e., three SNR levels and four methods plus the unprocessed case. Participants are introduced to the task in a training session with 4 words. The material is diotically presented using a pair of Roland RH-200S headphones. Listeners hear each word once in an arbitrary presentation order and are asked to indicate the word in a sheet list.

The intelligibility results for each method are presented in Fig. \ref{percep_test}. Each boxplot depicts the median and deviation values scores (\%) for one scenario, separating the (a) male, (b) female volunteers, and (c) the overall scores. The proposed method accomplishes intelligibility under all conditions over the competing approaches. For male listeners the HDAG obtained average intelligibility scores of 66\%, 85\% and 93\% compared to 52\%, 66\% and 86\% in the GTF$_\text{F0}$ technique for SNR values of -5 dB, 0 dB and 5 dB, respectively. Furthermore, female volunteers presented higher intelligibility rates than male, mainly for -5 dB with 75\% and 65\% for HDAG and GTF$_\text{F0}$. The overall results show again the superiority of HDAG with average scores of 71\%, 86\% and 92\%, surpassing GTF$_\text{F0}$ (59\%, 71\% and 86\%) and PACO (43\%, 64\% and 78\%). In accordance with findings in the objective measures ESTOI and ASII$_\text{ST}$, SSFV attains scores less or equal the UNP case.

\begin{table}[t!] \caption{\label{processing_time} Normalized Mean Processing Time.}
\renewcommand{\arraystretch}{1.5}
\begin{center}
{
\begin{tabular}{cccc}

\hline
SSFV&PACO&GTF$_\text{F0}$&HDAG\\\hline

0.32&0.67&0.89&1.00\\ \hline
\end{tabular}
}
\end{center}
\end{table}

\subsection{Normalized Processing Time}

Table \ref{processing_time} indicates the computational complexity which refers to the normalized processing time required for each method evaluated for 512 samples per frame. These values are obtained with an Intel (R) Core (TM) i7-9700 CPU, 8 GB RAM, and are normalized by the execution time of the proposed HDAG solution. The processing time required for F0 estimation and accurate harmonic adjust is also considered here. Note that the HDAG and GTF$_\text{F0}$ schemes present a longer processing time, since the FSFFE low/high pitch classification and HHT-Amp estimation are based on the EEMD, and demand a relevant computational cost.

\section{Conclusion}

This paper introduced the HDAG method for speech intelligibility enhancement in harmonic components of noisy speech. It is composed by four main steps. First, the HHT-Amp technique is adopted to estimate the F0 from voiced frames. The FSFFE separation was used for the detection and adjustment of these estimates, improving their accuracy. Then, selective Gammachirp filterbank was applied to the frames considering third-octave bands to best cover the regions most relevant to intelligibility. Finally, the filtered components were amplified by gain factors regulated by low/high pitch classification. Extensive experiments were conducted to evaluate the intelligibility enhancement provided by HDAG method and competitive approaches. Six acoustic noises were considered with four SNR values. Three objective measures are adopted for objective evaluation of speech intelligibility and quality. The results demonstrate that HDAG method outperformed the competitive approaches, with higher intelligibility and quality assessment in most noisy environments. A perceptual test for male and female listeners corroborated the objective results. Future research includes the investigation of the proposed method for other conditions, such as intelligibility enhancement for noisy reverberant speech.

\ifCLASSOPTIONcaptionsoff
  \newpage
\fi



%




\bibliographystyle{ieeetr}
\bibliography{tasl}

\end{document}